\documentclass[twocolumn,epjc3]{svjour3}          

\usepackage{graphicx}
\usepackage[utf8]{inputenc}
\usepackage{float}
\usepackage[english]{babel}

\usepackage{color}

\usepackage{amsmath}

\RequirePackage[T1]{fontenc}

\RequirePackage{graphicx}
\RequirePackage{mathptmx}      
\RequirePackage{flushend}
\RequirePackage[colorlinks,citecolor=blue,urlcolor=blue,linkcolor=blue]{hyperref}

\usepackage{sidecap}
\usepackage{caption}
\usepackage{subcaption}
\usepackage{amsmath}
\usepackage{braket}
\usepackage{multirow}
\usepackage{physics}

\journalname{Eur. Phys. J. C}

\begin{document}

\title{
Feasibility studies of the polarization of photons beyond the optical wavelength regime with the J-PET detector
}

\author{
P.~Moskal\thanksref{WFAIS}
\and
N.~Krawczyk\thanksref{WFAIS}
\and
B.~C.~Hiesmayr\thanksref{Vienna}
\and
M.~Ba{\l}a\thanksref{WFAIS}
\and    
C.~Curceanu\thanksref{LNF}
\and
E.~Czerwiński\thanksref{WFAIS} 
\and
K.~Dulski\thanksref{WFAIS} 
\and
A.~Gajos\thanksref{WFAIS} 
\and
M.~Gorgol\thanksref{UMCS} 
\and
R.~Del Grande\thanksref{LNF}
\and
B.~Jasi{\'n}ska\thanksref{UMCS}
\and
K.~Kacprzak\thanksref{WFAIS}
\and
L.~Kap{\l}on\thanksref{WFAIS}
\and
D.~Kisielewska\thanksref{WFAIS}
\and
K.~Klimaszewski\thanksref{SWIERK}
\and
G.~Korcyl\thanksref{WFAIS}
\and
P.~Kowalski\thanksref{SWIERK} 
\and    
T.~Kozik\thanksref{WFAIS} 
\and
W.~Krzemień\thanksref{SWIERKHEP}
\and
E.~Kubicz\thanksref{WFAIS}
\and
M.~Mohammed\thanksref{WFAIS,Mosul}
\and    
Sz.~Niedźwiecki\thanksref{WFAIS} 
\and
M.~Pa{\l}ka\thanksref{WFAIS}
\and
M.~Pawlik-Niedźwiecka\thanksref{WFAIS} 
\and
L.~Raczy{\'n}ski\thanksref{SWIERK} 
\and
J.~Raj\thanksref{WFAIS}
\and
Z.~Rudy\thanksref{WFAIS} 
\and
S.~Sharma\thanks{WFAIS}
\and
M.~Silarski\thanksref{WFAIS} 
\and
Shivani\thanksref{WFAIS}
\and
R.~Y.~Shopa\thanksref{SWIERK}
\and
M.~Skurzok\thanksref{WFAIS}
\and
W.~Wi{\'s}licki\thanksref{SWIERK}
\and
B.~Zgardzi{\'n}ska\thanksref{UMCS} 
}
\institute{Faculty of Physics, Astronomy and Applied Computer Science, Jagiellonian University,  S.~Łojasiewicza 11, 30-348 Kraków, Poland\label{WFAIS}
    \and
    University of Vienna, Faculty of Physics, Boltzmanngasse 5, 1090 Vienna, Austria\label{Vienna}
    \and
    INFN, Laboratori Nazionali di Frascati CP 13,  Via E. Fermi 40, 00044, Frascati, Italy\label{LNF}
    \and 
    Department of Nuclear Methods, Institute of Physics, Maria Curie-Sklodowska University, Pl.~M.~Curie-Sklodowskiej~1, 20-031 Lublin, Poland\label{UMCS}
    \and
    Świerk Computing Centre, National Centre for Nuclear Research,  05-400 Otwock-Świerk, Poland\label{SWIERK}
    \and
    High Energy Department, National Centre for Nuclear Research,  05-400 Otwock-Świerk, Poland\label{SWIERKHEP}
    \and
    Department of Physics, College of Education for Pure Sciences, University of Mosul, Mosul, Iraq\label{Mosul}
}
\maketitle
\begin{abstract}
J-PET is a detector optimized for registration of photons from the electron-positron annihilation via plastic scintillators where photons interact predominantly via Compton scattering. Registration of both primary and scattered photons enables to determinate the linear polarization of the primary photon on the event by event basis with a certain probability. Here we present quantitative results on the feasibility of such polarization measurements of photons from the decay of positronium with the J-PET and explore the physical limitations for the resolution of the polarization determination of 511keV photons via Compton scattering. For scattering angles of about 82$^\circ$ (where the best contrast for polarization measurement is theoretically predicted) we find that the single event resolution for the determination of the polarization is about 40$^\circ$ (predominantly due to properties of the Compton effect). However, for samples larger than ten thousand events the J-PET is capable of determining relative average polarization of these photons with the precision of about few degrees. The obtained results open new perspectives for studies of various physics phenomena such as quantum entanglement and tests of discrete symmetries in decays of positronium and extend the energy range of polarization measurements by five orders of magnitude beyond the optical wavelength regime.

\keywords{Positronium \and J-PET \and Compton scattering \and Polarization }
\end{abstract}

\section{Introduction}
Polarization is with no doubt one of the most interesting physical properties photons exhibit. It has been utilized to show several of the most basic foundations of quantum mechanics, where mainly optical (low energetic - few eV) photons were generated. However, thus far there were no studies where the degree of polarization was explored in measurements of high energy photons (in the range of MeV) originating from annihilations of positronium atoms. Polarization of such photons cannot be determined with optical methods. Here we show how it can be estimated via Compton scattering based on the well-known Klein-Nishina formula~\cite{Klein2013} and recent quantum information theoretical considerations~\cite{beatrix-arXiv:1807.04934}.

Measurement of the polarization degree of freedom of photons from positronium decay may open new possibilities in testing the discrete symmetries (T, CP and CPT symmetry) in leptonic sector since they provide a new class of operators~\cite{ACTA2016}. In addition, investigation of multi-partite entanglement of annihilation photons becomes possible~\cite{Beatrix-Science-Report2017,Nowakowski,Acin}.

The photon is a transverse electromagnetic wave and Compton scattering occurs most likely in the plane perpendicular to the electric vector of the photon~\cite{Klein2013,Evans1958}. Thus we can estimate the direction of its linear polarization $\epsilon$ by  the product of photons' momentum vectors before ($\hat{k}~=~\frac{\vec{k}}{|k|}$) and after ($\hat{k'}~=~\frac{\vec{k'}}{|k'|}$) the scattering~\cite{ACTA2016}, namely  $\hat{\epsilon} = \hat{k} \times \hat{k'}$. Note that we assumed here that the polarization vector is a real-dimensional vector, for more details in this Compton-context see Ref.~\cite{beatrix-arXiv:1807.04934}.

\begin{figure}[t]
  \includegraphics[width=0.5\textwidth]{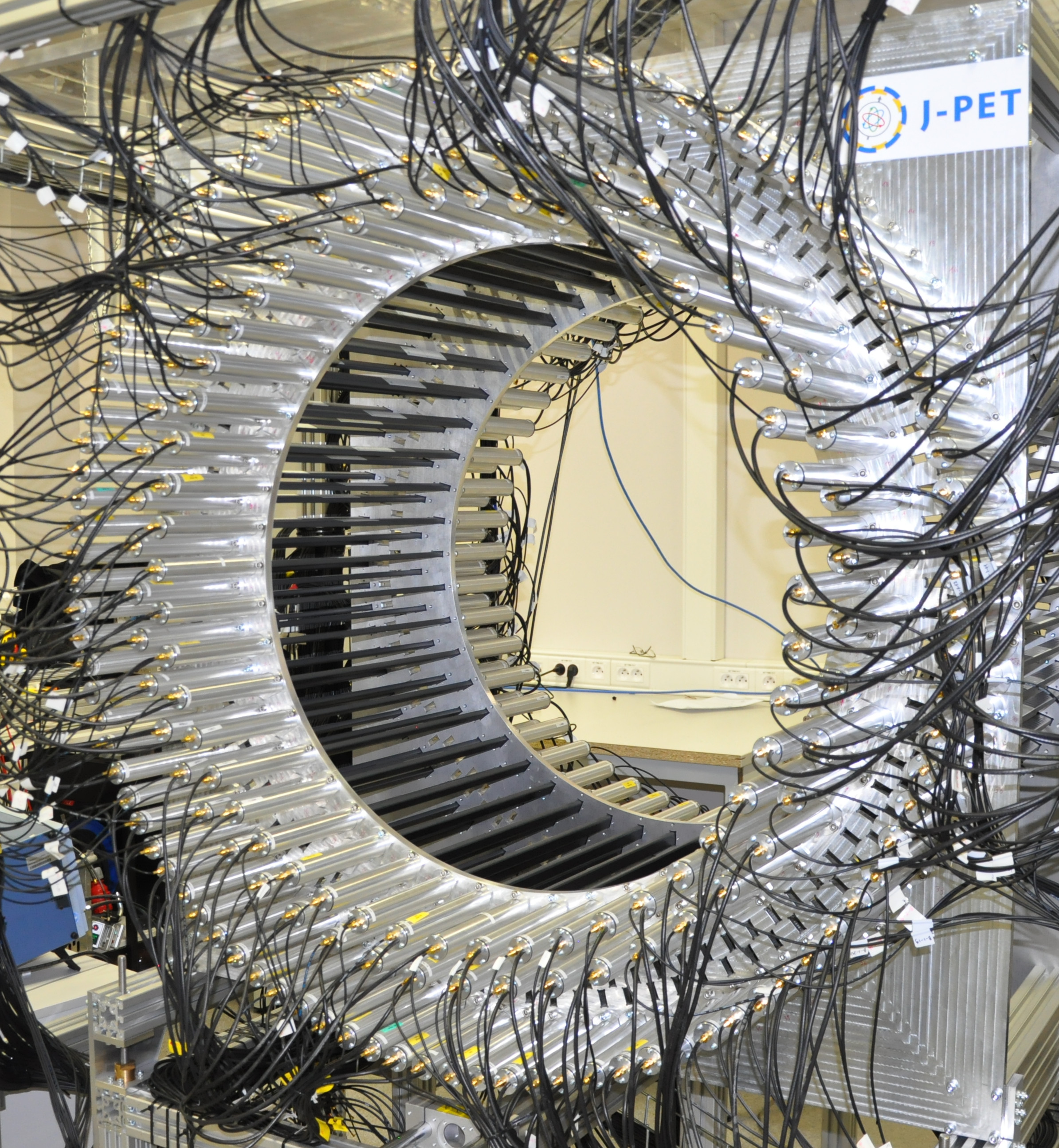} 
\caption{Photo of the J-PET detector. The inner part of the detector is of a cylindrical shape with the length of 
50~cm and diameter of 85~cm. The detector is made of three layers of plastic scintillator strips (black) and readout by vacuum tube photomultipliers (gray). 
\label{J-PETphoto}       
}
\end{figure}

The J-PET detector is built of plastic scintillator strips consisting mostly from carbon and hydrogen. Due to the low atomic number of these elements photons from the positronium annihilation interact in plastic scintillators predominantly via the Compton effect and a significant fraction of them may undergo two or even more subsequent scatterings in different strips. A picture of the present prototype of the detector is shown in Fig.~\ref{J-PETphoto}. Its geometry and properties are described in details in the references~\cite{NIM2014,NIM2015,PMB2016,ACTA2017}.
Therefore, here for completeness we mention only briefly its main characteristics. 
\begin{figure}[t] 
  \includegraphics[width=0.5\textwidth]{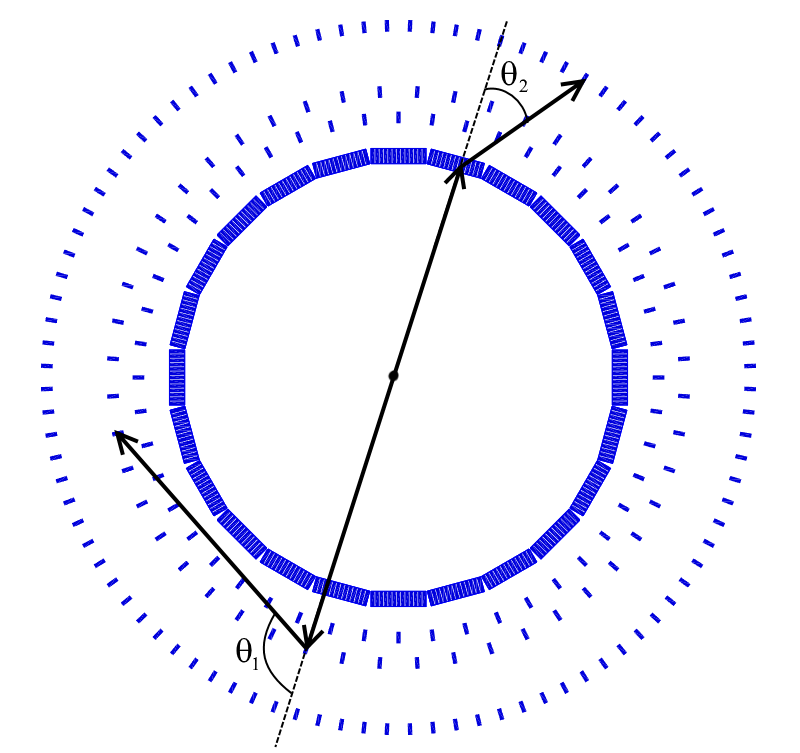}
\caption{Cross section of the updated J-PET detector. Superimposed arrows indicate primary and scattered momenta of photons originating from para-positronium decaying in the center of the detector.  Detector is build from plastic scintillators strips arranged axially in four rings with radii of $362$~mm, $425$~mm, $467.5$~mm, $575$~mm.
\label{DetectorSchemaSingle} 
}
\end{figure}
J-PET at the present stage is built from three cylindrical layers including in total $192$ plastic scintillators strips with dimensions of $7\times 19 \times 500~\textrm{mm}^3$. Light signals from each strip are converted to electrical signals by photomultipliers placed at opposite ends of the strip~\cite{NIM2014}. The position and time of the photons interacting in the detector material are determined based on the arrival time of light signals at both ends of the scintillator strips. The signals are probed in the voltage domain with the accuracy of about $30$ps by a newly developed type of front-end electronics~\cite{Paka2017} and the data are collected by the novel trigger-less and reconfigurable data acquisition system~\cite{Korcyl2016,Korcyl2018}. 
For the data processing and simulations a dedicated software framework was developed~\cite{Krzemien2015,Krzemien2015framework,Krzemien2016}. The hit-position and hit-time are reconstructed by the dedicated reconstruction methods based on the compressing sensing theory and the library of synchronized model signals~\cite{Raczynski2014,Raczynski2015,Raczynski2017,NIM2015}. Though the photons interact in the plastic scintillators predominantly via the Compton effect, the J-PET detector enables studies of positronium decays with angular resolutions of about~1$^\circ$~\cite{Daria-EPJ-2016}. Presently a new innermost layer is being installed and commissioned. This fourth layer consists of $312$ plastic scintillator strips with dimensions of $6\times 24 \times 500~\mbox{mm}^3$ read out by matrices of silicon photomultipliers, which is expected to triple the efficiency for the single photon detection and improve the time resolution by about a factor of $1.5$~\cite{PMB2016}.
Future measurements will be carried out with the full equipped detector, therefore for simulations conducted in this article we assumed the full version, i.e. a four layer geometry whose cross section is shown in Fig.~\ref{DetectorSchemaSingle}.

In this article 
we explore first the possibility of the determination of the polarization of annihilation photons in the case of an ideal detector system. Section~\ref{section2} provides an estimate of the accuracy of polarization determination as a function of the scattering angle for  $511$~keV--photons originating from the $e^+e^-$ annihilations into two photons.
Subsequently, in Section~\ref{section3} the capability of the determination of the relative angle between the polarization directions of a photon pair 
originating from the para-positronium decay $\mbox{p-Ps} \to 2\gamma$ is provided. Next, in Section~\ref{section4} the efficiency and angular resolution of the J-PET detector for studies of the relative polarizations angle for 
photons from positronium decay is presented. Finally, the obtained results and their implications for studies of quantum entanglement and discrete symmetries are summarized in Section~\ref{section5}.

\section{Determination of a single photon polarization via Compton scattering}
\label{section2}

\begin{figure}[h]
  \includegraphics[width=0.5\textwidth]{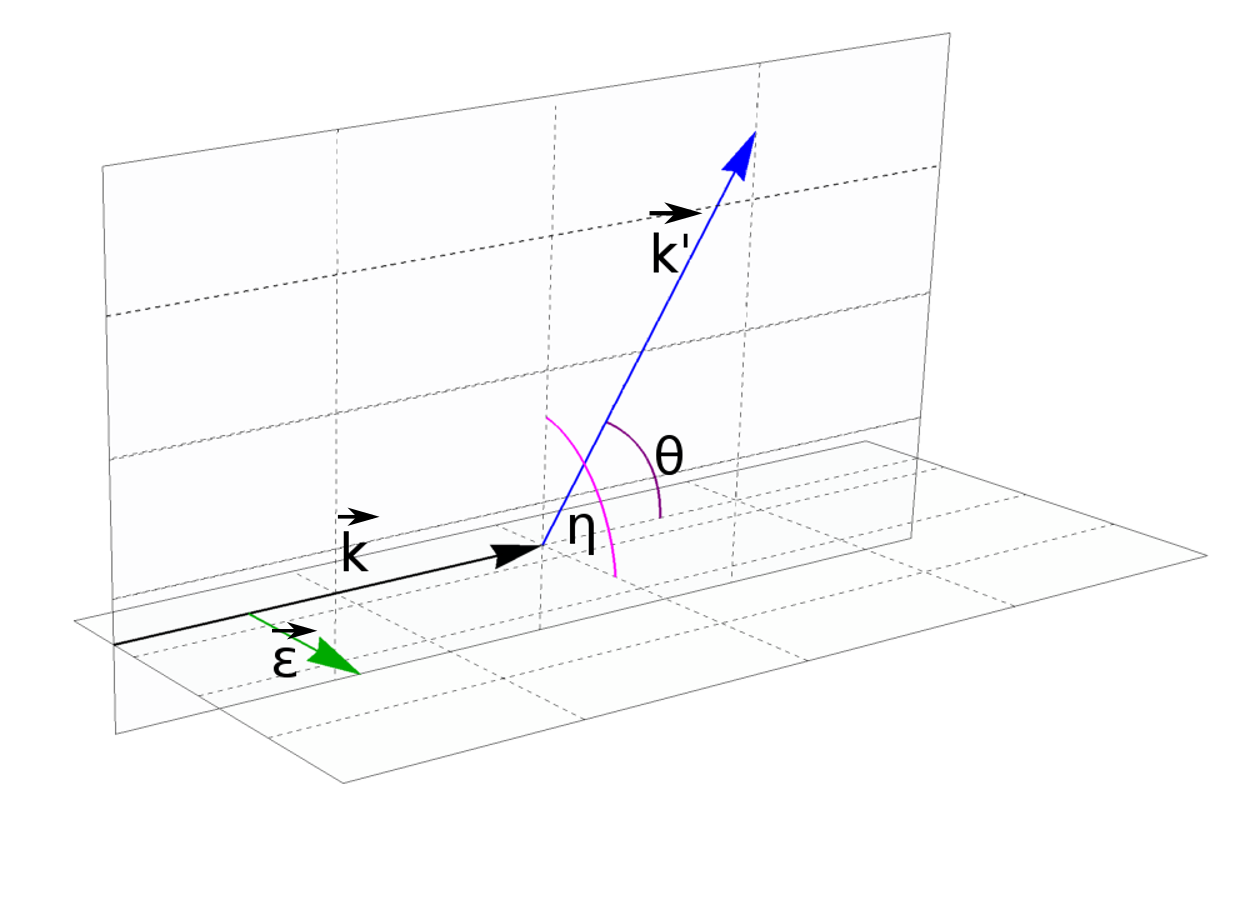}
\caption{Scheme of Compton scattering. An incident photon with momentum $\vec{k}$ scatters on an electron resulting in a change of its momentum $\vec{k^{\prime}}$. The Compton scattering angle is denoted by $\theta$. Initial and scattered momenta form a plane, which will be referred to as \textit{scattering plane}. The incident photon momentum $\vec{k}$ and its linear polarization vector $\vec{\epsilon}$ form a second plane, referred to as the \textit{polarization plane}. The angle between those two planes is denoted as $\eta$.
\label{fig:ComptonSchema}       
}
\end{figure}

\begin{figure}[h]
  \includegraphics[width=0.5\textwidth]{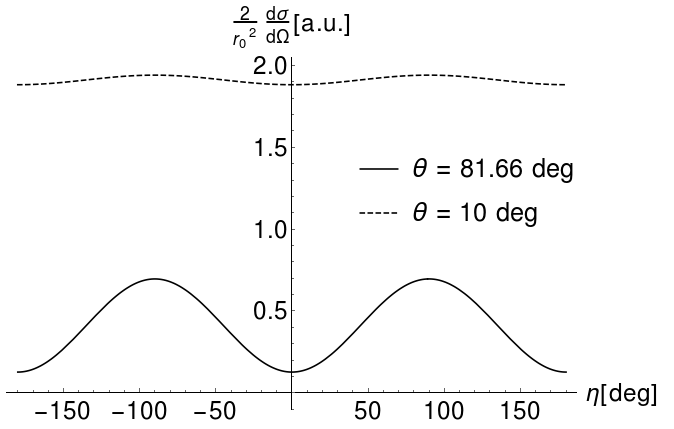}
\caption{Klein-Nishina differential cross section. The plots show the Klein-Nishina relation~(\ref{eq:P}) for photons with initial energy $E=511$~keV and scattered at angle $\theta = 81.66^\circ$ (solid line) and $\theta = 10^\circ$ (dashed line) as function of the angle $\eta$.
\label{fig:KleinNishina} 
}
\end{figure}

Angular distributions of photons scattered on an electron are described by the Klein-Nishina differential cross section~\cite{Klein2013,Evans1958}
\begin{eqnarray}
\frac{d\sigma (E, \theta, \eta)}{d\Omega} &=& \frac{r_{0}^2}{2} \left(\frac{E^{\prime}}{E}\right)^2 \left(\frac{E}{E^{\prime}} + \frac{E^{\prime}}{E} - 2\sin^2{\theta}\cos^2{\eta}\right) 
\label{eq:KNcross}
\end{eqnarray}
with
\begin{eqnarray}\label{energydep}
 E^{\prime}(E, \theta) &=& \frac{E}{1 + \frac{E}{m_{e}c^2}(1 - cos{\theta})}\;,
\end{eqnarray}
where $E$ is the energy of initial photon, $E^{\prime}$ is the energy of photon after scattering, $\theta$ is the Compton scattering angle and $\eta$ is the angle between scattering and polarization planes (for definition see also Fig.~\ref{fig:ComptonSchema}). 
There are two important limits that can be seen from Eq.~(\ref{eq:KNcross}). If the scattering angle $\theta$ is close to zero or to 180$^\circ$, the variation of cross section with $\eta$ is not observable. A scattering at $\eta~=~90^\circ$ makes the last term maximal, however, the energy of the outgoing photon depends also on the Compton scattering angle $\theta$, (Eq.~(\ref{energydep})). Thus the visibility, the interference contrast of the oscillation in $\eta$, the angle between the scattering and polarization planes, is a function of energy and scattering angle, i.e. \begin{eqnarray}\mathcal{V}(\theta,E_i)
&:=&\frac{\max_\eta\{\frac{d\sigma (E, \theta, \eta)}{d\Omega}\}-\min_\eta\{\frac{d\sigma (E, \theta, \eta)}{d\Omega}\}}{\max_\eta\{\frac{d\sigma (E, \theta, \eta)}{d\Omega}\}+\min_\eta\{\frac{d\sigma (E, \theta, \eta)}{d\Omega}\}}\nonumber\\
&=&\frac{\sin^2\theta}{\frac{E}{E^{\prime}} + \frac{E^{\prime}}{E} -\sin^2\theta}\;.
\label{eq:visibility}
\end{eqnarray}
For $511$~keV--photons the optimal $\theta$ equals to $81.66^\circ$, i.e. gives the maximal visibility for the variation of the azimuthal angle $\eta$. In Fig.~\ref{fig:KleinNishina} we have plotted therefore the cross section for two cases:  for arbitrarily chosen small angle $\theta=10^\circ$ and the optimal angle $\theta=81.66^\circ$. From the Klein-Nishina formula~\eqref{eq:KNcross} we deduce directly that the scattering cross section favors small Compton scattering angles $\theta$ over large ones (exemplified also in Fig.~\ref{fig:KleinNishina}).
Furthermore, we observe an azimuthal asymmetry $\eta$ of the scattered radiation (see Fig.~\ref{fig:KleinNishina}), namely 
we find that scatterings around $\eta = \pm 90^\circ$ are favored over those of $\eta=0^\circ$ and $\eta=180^\circ$. This finds a simple physical interpretation: since an initial polarization vector component normal to the scattering plane does not need to change its orientation for the outgoing photon since it is still normal to the new momentum vector. Whereas, a polarization vector component in the scattering plane has to change by $\cos\theta$ since the polarization vector needs to be normal to the new momentum vector.
\begin{figure}[h]
  \includegraphics[width=0.5\textwidth]{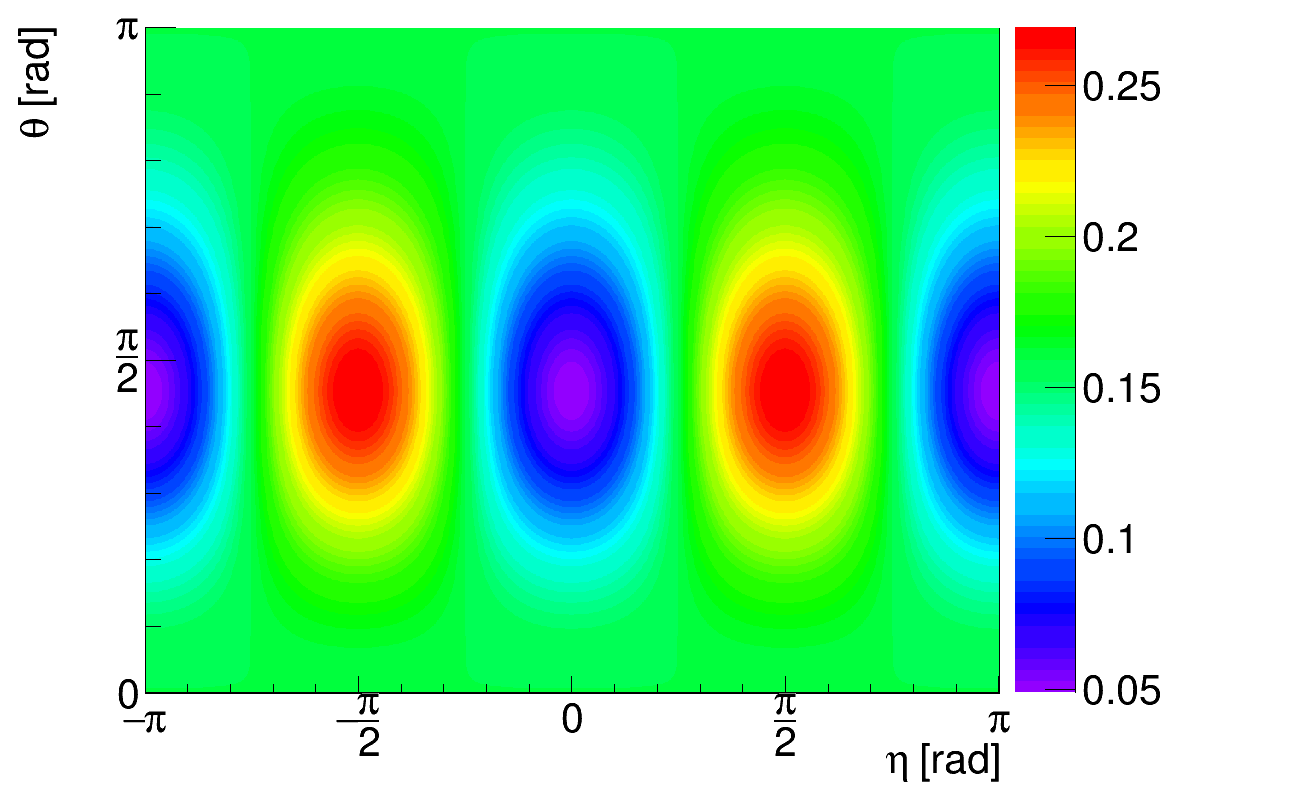}
\caption{Normalized Klein-Nishina
relation~(\ref{eq:P}). In order to compare distributions of azimuthal angle $\eta$ for radiation scattered under different Compton angles we need to perform a proper normalization. For each scattering angle $\theta$, corresponding to horizontal lines in the plot, we have normalized the probability as function of $\eta$. Around $\theta=81.66^\circ$ the biggest variation is expected.}
\label{fig:NormKlein}   
\end{figure}

Due to different values of the cross section for photons scattered under various Compton angles $\theta$, in order to compare the likelihoods of scattering parallel and normal to polarization, we introduce the following normalization $P(E, \theta, \eta)$, which under fixed initial energy $E$ and Compton scattering angle $\theta$ can be understood as the probability density distribution of the angle $\eta$:
\begin{equation}
P(E, \theta, \eta) = N(E, \theta) \cdot \frac{d\sigma(E, \theta, \eta)}{d\Omega}\; ,
\label{eq:P}
\end{equation}
where the normalization factor $N(E, \theta)$ is defined as:
\begin{equation}
N(E, \theta) = 1\slash\int\limits_{-\pi}^{\pi}\frac{d\sigma(E, \theta, \eta)}{d\Omega}  d\eta\;.
\end{equation}

Fig.~\ref{fig:KleinNishina} and Fig.~\ref{fig:NormKlein} clearly show that indeed for the scattering angles around $\theta~=~82^\circ$ the probability of the scattering has its maximum value when the scattering plane is perpendicular to the direction of the electric vector of the primary photon ($\eta = \pm 90^\circ$). It is important to stress that it is the case independently of the scattering angle $\theta$, even though for small angles the maximum is hardly visible.
This observation implies that, as stated in the introduction, we can estimate the direction of the linear polarization of the Compton scattering photon by constructing a product of the momentum vectors of the photon between and after the scattering: 
$\hat{\epsilon} = \hat{k} \times \hat{k'}$ ~\cite{ACTA2016}. 
When using such a definition of $\epsilon$, we may interpret the normalized Klein-Nishina differential cross section as a probability density distribution of the deviation between 90$^\circ$ and the real angle $\eta$. Thus, for example the solid curve in Fig.~\ref{fig:KleinNishina}, after normalization to unity could be seen as a physical limitation of the
achievable resolution for the determination of the direction of the polarization
of 511~keV photons scattered under  $\theta~=~81.66^\circ$.
\begin{figure}[h]
 \includegraphics[width=0.5\textwidth]{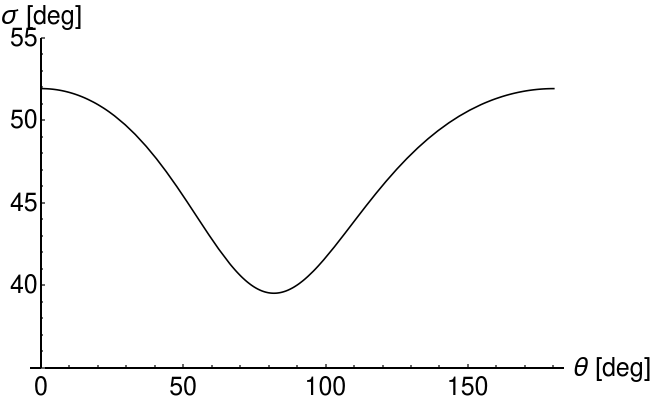}
\caption{Standard deviation $\sigma$ of the probability density distribution (\ref{eq:P}) as a function of $\theta$ for $511$~keV--photons. The minimum is $40^\circ$ for a scattering angle of $81.66^\circ$ and the maximum ($52^\circ$) for small or large scattering angles.
\label{std}       
}
\end{figure}
Fig.~\ref{std} presents the physical limit for the resolution of determining the polarization direction for $511$~keV--photons when identifying above defined $\epsilon$  with the direction of photon's polarization. The obtained values of $\sigma$ were determined as a fit of a Gaussian function to the distributions of $\frac{d\sigma}{d\Omega}(\eta)$ in the range of positive values of $\eta$. As mentioned earlier, for  forward ($\theta=0^\circ$) and backward ($\theta=180^\circ$) scattering  the polarization direction cannot be determined via Compton scattering, which results in a standard deviation $\sigma$ close to $52^\circ$ as expected for the uniform distribution.

\section{Relative polarization of photons from positronium decay into 2$\gamma$}
\label{section3}

In the previous section we discussed the limitations of determining the direction of single photon polarization with respect to the scattering plane, when the measurement is based on the Compton scattering formula. In this section we will extend this discussion to explore the possibilities of measurements of the relative angle between polarization directions of maximally entangled photon pairs originating from the decay of para-positronium. 
\begin{figure}[h]
  \includegraphics[width=0.5\textwidth]{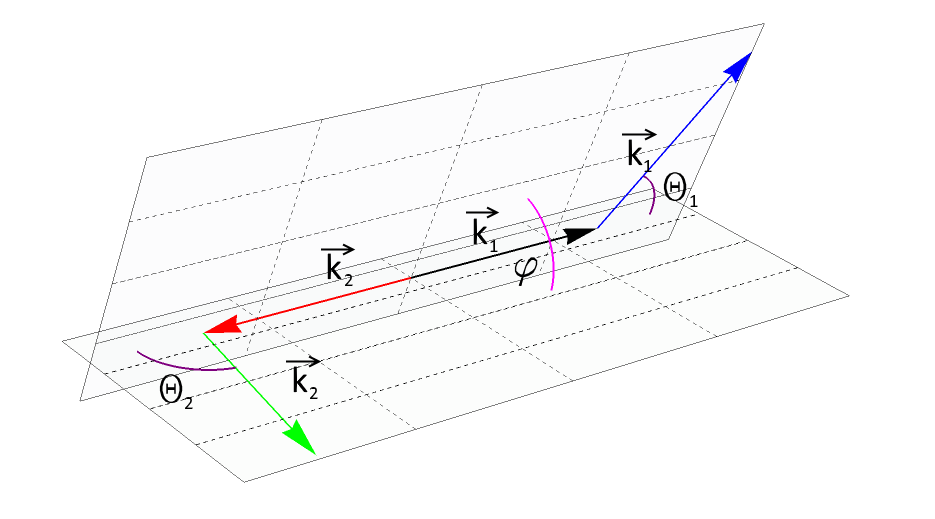}
\caption{Schematic description of Compton scattering of two gamma photons originating from the positronium annihilation. Angle between scattering planes is denoted as $\vec{\varphi}$.
\label{fig:BackToBack}  
}
\end{figure}
The Bose-symmetry and the parity conservation in the decay of para-positronium (for details see ~\cite{beatrix-arXiv:1807.04934,Harpen2003}) imply that
the state of the resulting two photons is maximally entangled. In the linear polarization basis ($\ket{H},\ket{V}$) 
with respect to one fixed coordinate system, the state can be written as
\begin{equation}
\ket{\psi} = \frac{1}{\sqrt{2}}\left\{\ket{H}_1\otimes\ket{V}_2 + \ket{V}_1\otimes\ket{H}_2\right\}\;,
\label{eq:psi} 
\end{equation}
where $\ket{H}$ and $\ket{V}$ denote the corresponding horizontal and vertical polarized states perpendicular to the photons propagation direction. It is important to note, as indicated pictorially in Fig.~\ref{fig:BackToBack}, that both photons originating from the decay of para-positronium propagate in its rest-frame along the same axis. Equation~\eqref{eq:psi} implies in addition that the polarization states of photons $1$ and $2$ are orthogonal to each other.

Thus we have to simulate events of two emitted photons assuming that for each event at the moment of Compton scattering (the measurement act) the  relative angle between the polarization directions of photons $1$ and $2$ is equal to $90^\circ$. Our overarching goal is to obtain the feasibility of deducing the correlations with the J-PET setup, therefore we do not invoke the predicted entanglement by simulating directly the joint scattering cross section, e.g. given in Ref.~\cite{beatrix-arXiv:1807.04934}, but simulate the separable states, $\ket{HV}$ and $\ket{VH}$. Herewith, the theoretical predicted uncertainties of the Compton scattering process are taken into account (our goal) without invoking the theory based on the quantum numbers in the decay of the positronium (except orthogonal polarisation in the moment of scattering). Note that as discussed in details in Ref.~\cite{beatrix-arXiv:1807.04934} the entanglement would be recognizable experimentally by observation in mutually unbiased bases/settings, revealing the stronger correlations exhibited by entangled states compared to separable states. Moreover, as we outline later our final simulations differ purely by a factor that can be easily inserted to the final result.

In the previous section, it was shown that the polarization direction $\hat{\epsilon}$ of a single photon can be estimated as a direction perpendicular to the scattering plane. Therefore, the relative angle between the polarization direction estimators ($\angle(\hat\epsilon_1,\hat\epsilon_2)$) is equal to the angle  between scattering planes, denoted by $\varphi$ in Fig.~\ref{fig:BackToBack}. Thus, this angle $\varphi$ may be treated as an estimator of the relative polarization directions when measured via Compton scattering. Every single measurement is limited by the resolution described by the Klein-Nishina formula~\ref{eq:KNcross} (as discussed  in detail in the previous sections). 

\begin{figure}[H]
  \includegraphics[width=0.5\textwidth]{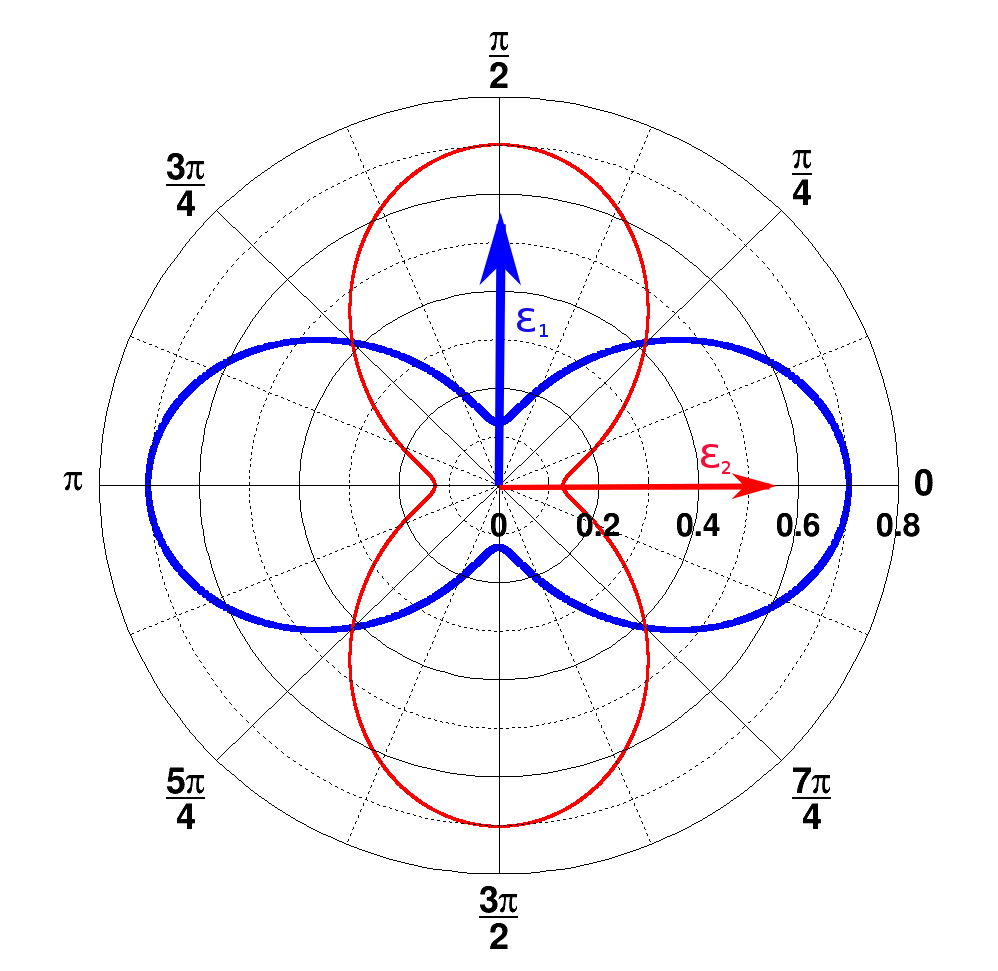}
  \includegraphics[width=0.5\textwidth]{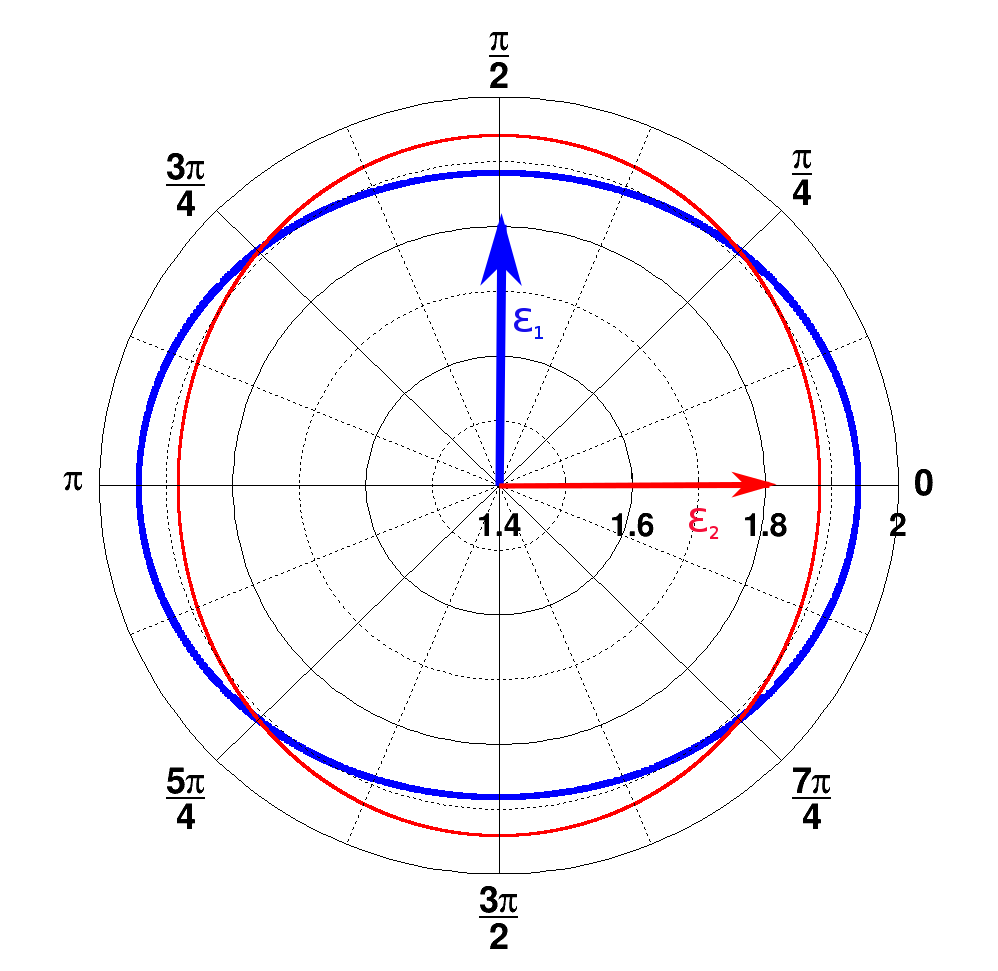}
\caption{Radial plots of Klein-Nishina cross sections for $511$~keV photons with  polarization vectors $\vec{\epsilon_1}$ (thick blue line) and $\vec{\epsilon_2}$ (thin red line), orthogonal to each other. Values given in radians at the plot indicate angle $\eta$. The values of cross sections for a given $\eta$ angle correspond to the distance from the center of the plot to the point on the line. The upper panel shows results for Compton scattering angle of $\theta_1~=~\theta_2~=~ 81.66^\circ$ and the lower for $\theta_1~=~\theta_2~=~10^\circ$. 
\label{fig:polar81} 
}
\end{figure}

In Fig.~\ref{fig:polar81} we present the distributions of Klein-Nishina cross sections for two orthogonal polarized photons in the form of radial plots. Upper plot shows the result for the case when both scatterings occurred under $81.66^\circ$ (best resolution for the polarization determination). 
The lower panel presents results for the case when \\ $\theta_1~=~\theta_2~=~10^\circ$, where the resolution of determining $\eta$ angle is much lower resulting in the nearly overlapping curves representing the two possible polarization states.

\begin{figure}[h]
  \includegraphics[width=0.5\textwidth]{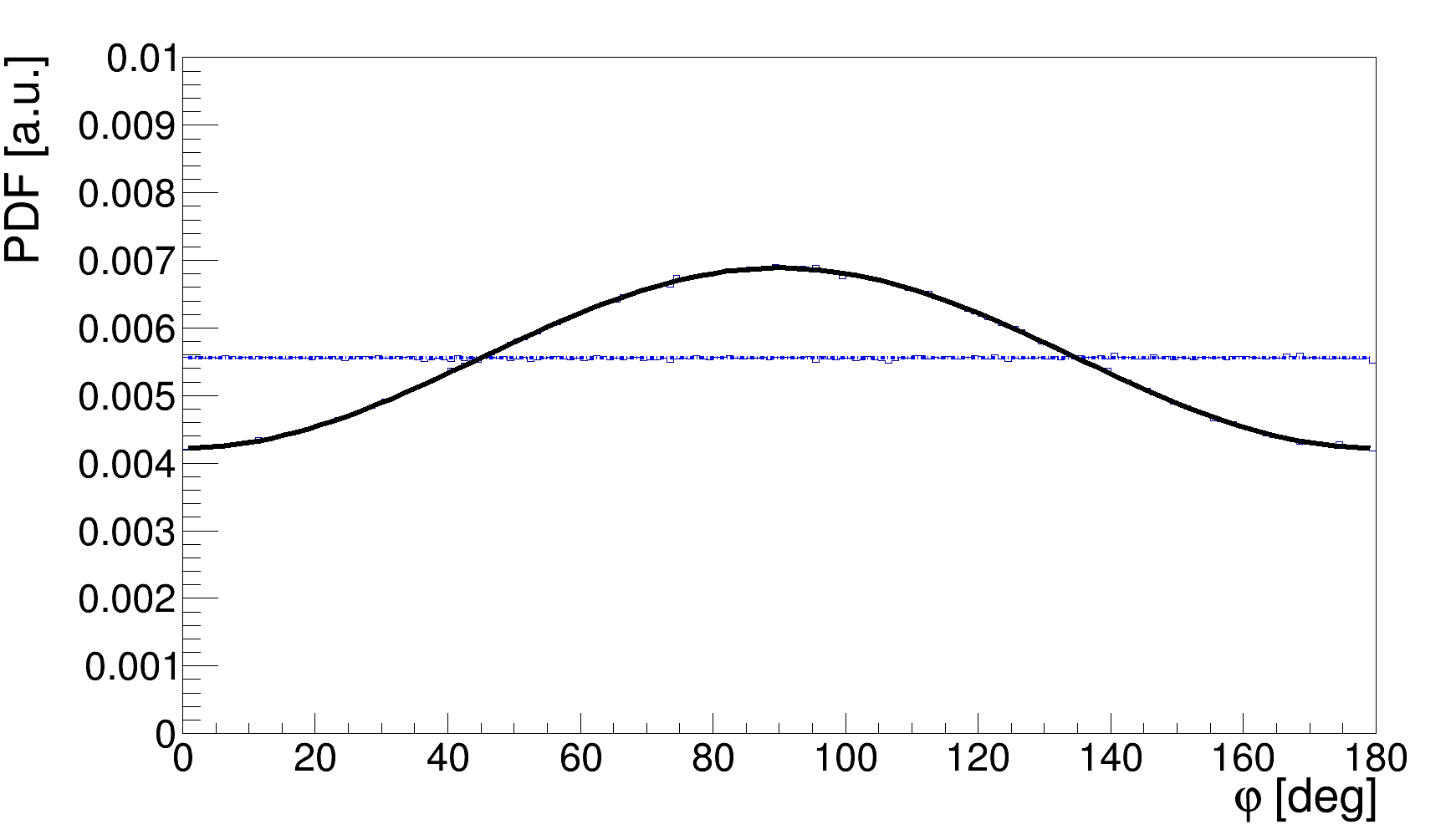}
\caption{Distribution of the angle $\varphi$ between scattering planes of the annihilation photons. The black continuous line corresponds to both photons scattered at Compton angle equal to $81.66^\circ$, while the blue dashed line to the case where both photons scattered at an angle of $10^\circ$. The simulated histograms are superimposed with the result of the fit of  function (\ref{cosine}) (for details see text). The simulated histograms and results of the fit are overlapping each other.
\label{TheoSim}
}
\end{figure}

In order to quantify the effect we have performed Monte-Carlo simulations of $\mbox{p-Ps} \to 2\gamma$ events, assuming that polarizations of photons in each event are orthogonal and generating for each photon independently an angle $\eta$ according to the Klein-Nishina distribution~\eqref{eq:KNcross}. Next, for each event a relative angle between scattering planes ($\varphi$) was calculated.
The distribution of this angle $\varphi$, which we treat as an estimator of the measured relative angle between the polarization directions,
is presented in Fig.~\ref{TheoSim}. The solid line shows the result for the case of the highest visibility $\mathcal{V}$, Eq.~(\ref{eq:visibility}), at $\theta_1~=~\theta_2~=~81.66^\circ$ and the dashed line for the $\theta_1~=~\theta_2~=~10^\circ$ (close to zero visibility). These studies imply that the determination of the polarization degrees of freedom in the decays of positronium, even assuming ideal  detectors, will be only effective for scatterings angles $\theta$ close to the values of $82^\circ$.

Here in order to calculate the visibility as a function of scatterings angles, for each pair of $\theta_1,\theta_2$ (within $1^\circ$ grid) we have simulated $\varphi$ distribution and fitted the formula:
\begin{equation}
f(\varphi) = A\cdot \cos^{2}{(\varphi - \delta)} + B,
\label{cosine}
\end{equation}
which describes very well the simulated $\varphi$ distributions with $A$, $B$ and $\delta$ being free parameters of the fit. One exemplary result of the fit is shown in Fig.~\ref{TheoSim} and shows that the theory predictions overlap well with the simulation. Based on the equations, (\ref{cosine}) and~(\ref{eq:visibility}), the visibility squared is calculated as $\mathcal{V}^2~=~\frac{A}{2B+A}$.
\begin{figure}[H]
  \includegraphics[width=0.5\textwidth]{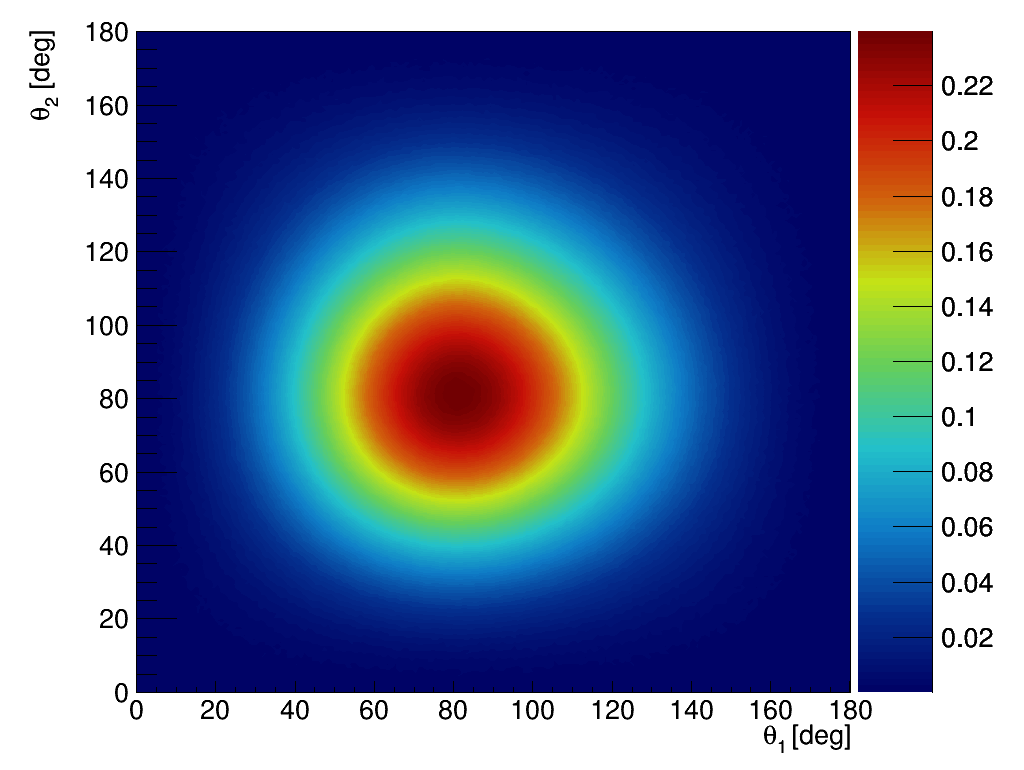}
\caption{Visibility$\mathcal{V}^2$ as a function of the scattering angles $\theta_1$ and $\theta_2$.  
\label{fig:visibility}
}
\end{figure}
A contour plot of $\mathcal{V}^2$ is given in Fig.~\ref{fig:visibility} and it shows that in case of back-to-back $511$~keV--photons, in order to measure effectively the angle between their relative polarization directions, the detector should be designed in a way of maximizing efficiency for the scatterings angles close to $82^\circ$.

\section{Feasibility of $Ps\rightarrow 2\gamma$ studies with J-PET}
\label{section4}

\begin{figure}[h]
  \includegraphics[width=0.5\textwidth]{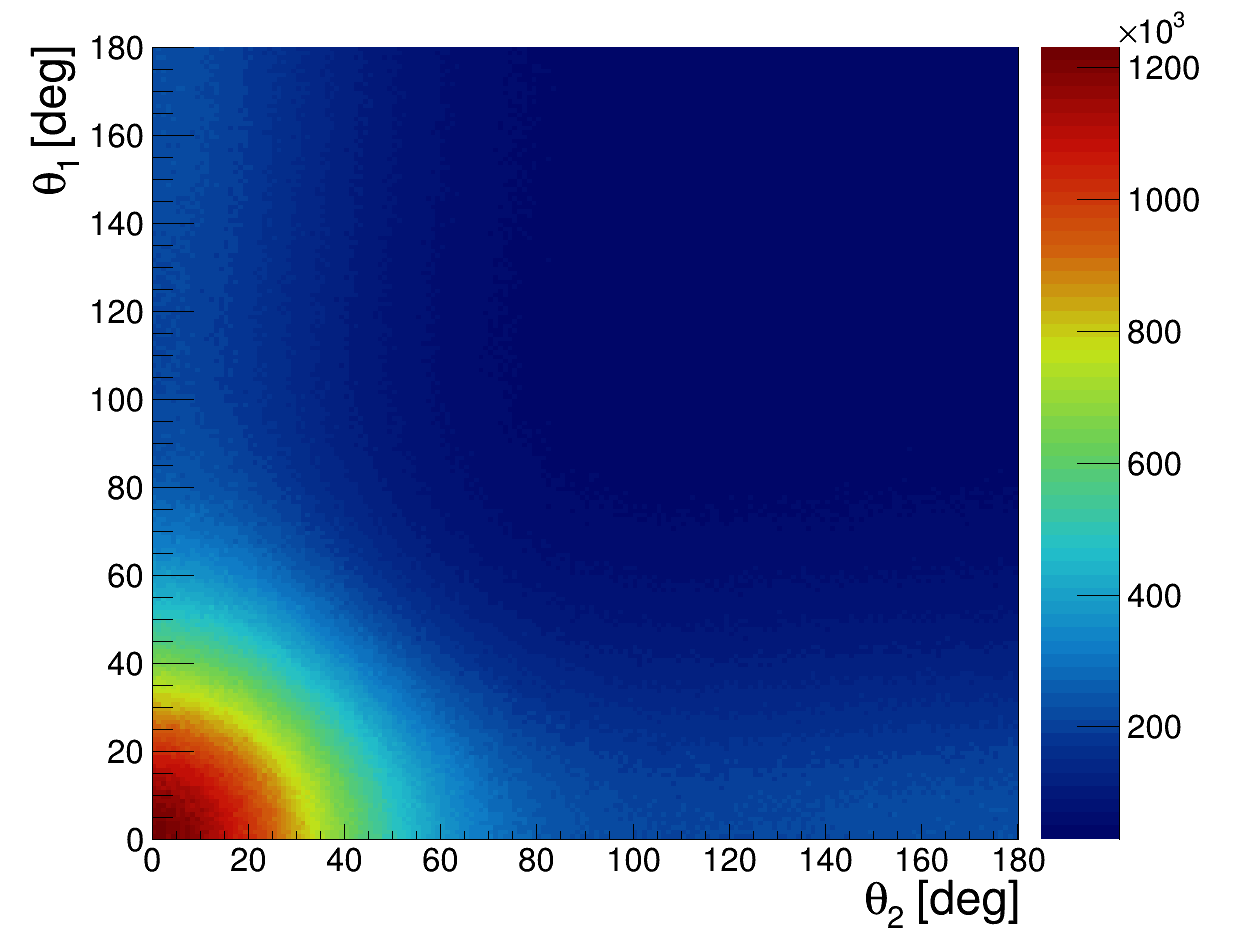}
\caption{Event distribution as a function of Compton scatterings angles $\theta_1$ and $\theta_2$ simulated for the case of an ideal detector with 100\% registration efficiency. The figure is based on $5.96 \times 10^9$ simulated events.
\label{fig:Theory-Theta1-vs-Theta2}
}
\end{figure}

\begin{figure}[h]
  \includegraphics[width=0.5\textwidth]{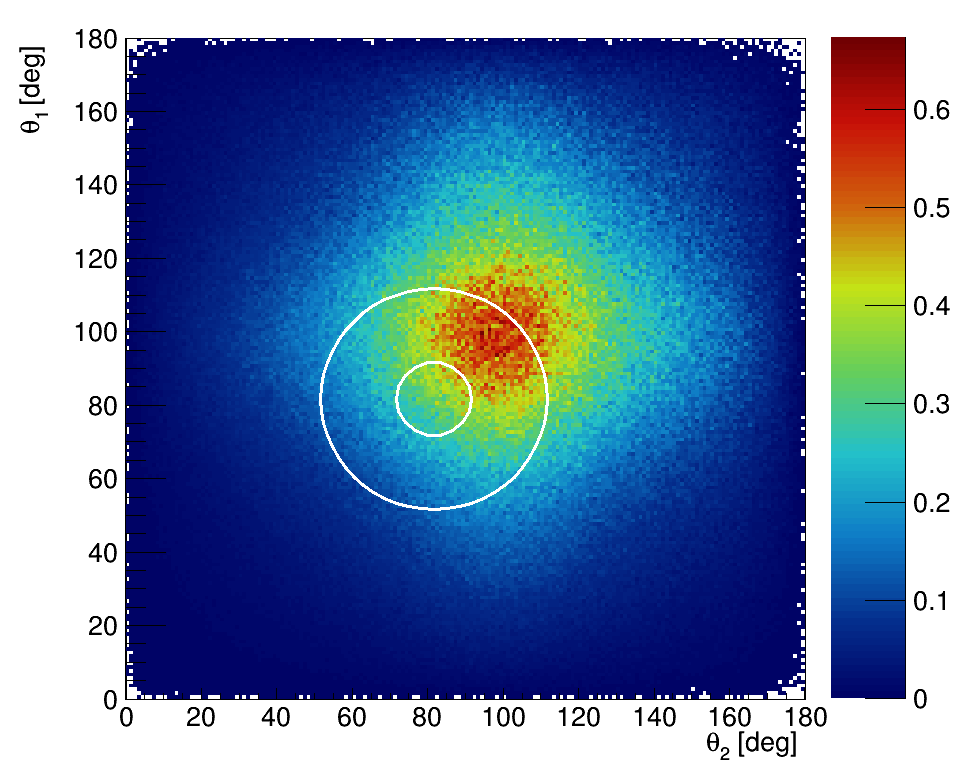}
\caption{Idealized (not including event selection criteria) detection efficiency of the J-PET detector as a function of $\theta_1$ vs $\theta_1$. The superimposed circles indicate regions  with radius of R~=~10$^\circ$ and R~=~30$^\circ$ around the point of the highest visibility $\theta_1~=~\theta_2~=~81.66^\circ$.
\label{fig:Efficiency-Theta1-vs-Theta2}
}
\end{figure}

In order to study the feasibility of the measurement of $\mbox{Ps}\rightarrow 2\gamma$ with the subsequent Compton scattering of both gamma photons, as indicated in Fig.~\ref{DetectorSchemaSingle}, we first simulated the distribution of scattering angles $\theta_1$ versus $\theta_2$ in the case of the ideal detector, assuming that each event is  measured with the efficiency of 100\%.  The scattering angles $\theta_1$ and $\theta_2$ were generated independently with the probability density distribution corresponding to the Klein-Nishina formula~(\ref{eq:KNcross}). The result is shown in Fig.~\ref{fig:Theory-Theta1-vs-Theta2}. As expected it is strongly picked at forward angles, where the visibility $\mathcal{V}^2$ (shown in Fig.~\ref{fig:visibility}) is negligible.   However,  though the overarching aim of the \mbox{J-PET} detector was medical imaging~\cite{NIM2014,NIM2015,PMB2016,ACTA2017,PMB2018}, its  idealized efficiency (not including suppression due to the hardware thresholds and software selection criteria) for the registration of secondary scattered photons is maximized quite close to the region of the highest visibility (Fig.~\ref{fig:Efficiency-Theta1-vs-Theta2}). 

It is important to emphasize that events corresponding to a given pair of scattering angles ($\theta_1, \theta_2$) are registered by many different combinations of the scintillator strips. Due to the axial symmetry of the detector, all strips in the same layer contribute equally to a given bin in the  ($\theta_1, \theta_2$) plot. This suppresses systematical errors due to the uncertainty in the detectors dimensions and geometrical misalignment.

The simulations were performed using the GATE package. GATE (Geant4 Application for Tomographic Emission) is a Monte Carlo simulation platform developed by the OpenGATE collaboration \cite{Santin2003,Jan2004,Jan2011} based on Geant4 software. It is dedicated to numerical simulations in medical imaging and radiotherapy. It utilizes an easy macro mechanism to configure experimental settings for Computed Tomography,  Single Photon Emission Computed Tomography, Positron Emission Tomography as well as Optical Imaging (Bioluminescence and Fluorescence) or Radiotherapy.

\begin{figure}[h]
  \includegraphics[width=0.5\textwidth]{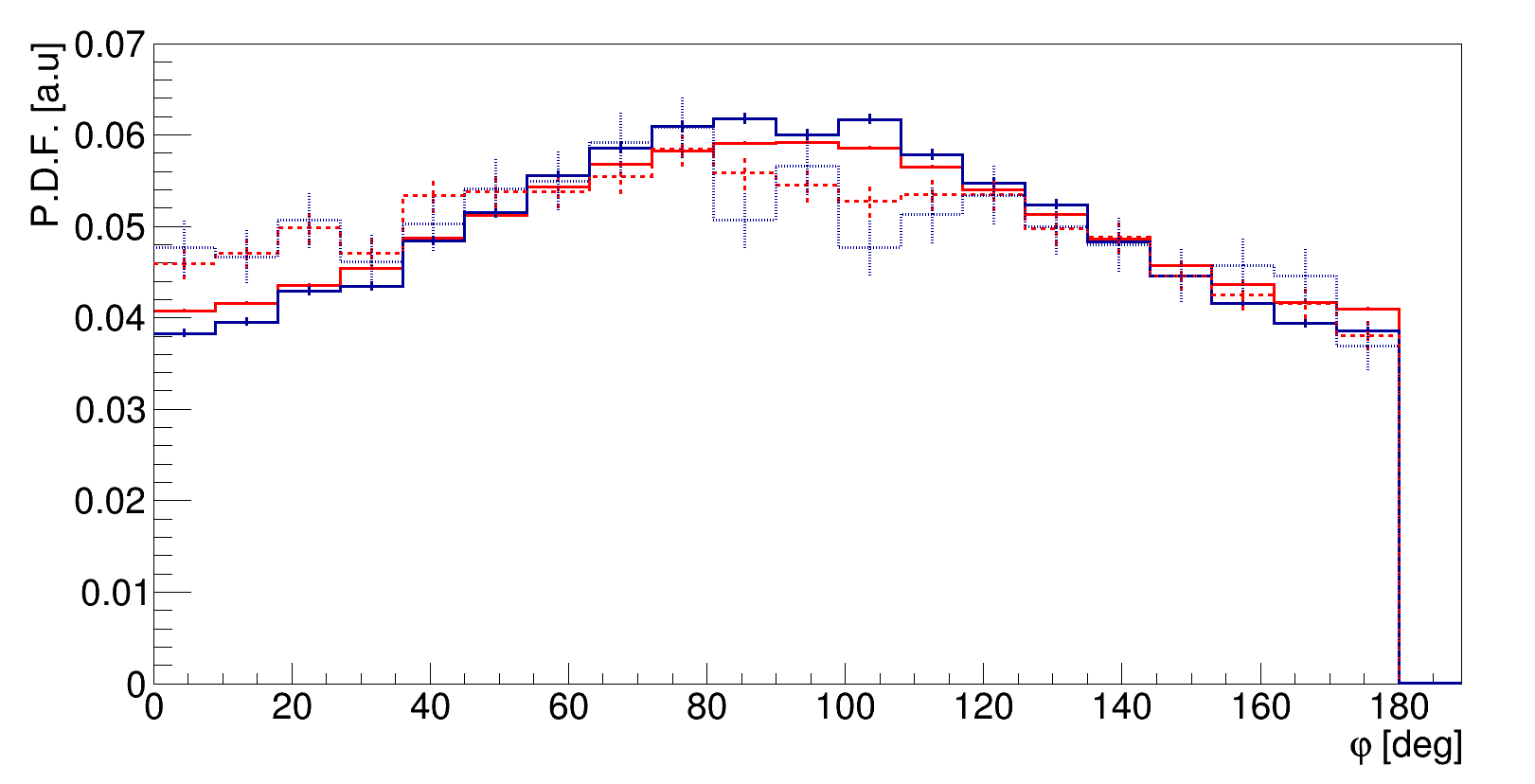}
\caption{Distribution of relative angle $\varphi$ between Compton scattering planes of photons from the $\mbox{p-Ps} \to 2\gamma$ process.  Black and red solid lines show results obtained for regions of high visibility for R~=~$10^\circ$ (black line) and R~=~$30^\circ$ (red line), respectively.
Dashed and dotted red lines indicate results as determined for signals simulated in the J-PET detector with the GATE package for R~=~$30^\circ$ and $d~>~12$~cm (dashed line) and for R~=~$30^\circ$, $d~>~12$~cm and energy loss larger than $50$~keV (dotted line).
\label{expected}
}
\end{figure}

In the simulations the full geometry of the J-PET detector and the composition of the detector material were taken into account. The interactions of gamma photons in the scintillators were simulated by GATE which uses Klein-Nishina formula~(\ref{eq:KNcross}). In the simulations we assumed that the source of positronium atoms is placed in the center of the detector and that the back-to-back gamma photons (each with energy of $511$~keV) from the $Ps\to 2\gamma$ annihilation are isotropically emitted.
The relative angle between the polarizations of the two photons (at the moment of interaction) was fixed to $90^\circ$, while the polarization direction of the single photon was distributed isotropically around the axis of photons propagation. Note that direction of the propagation axis varies from event to event, however, the axes are isotropically distributed.  
The histograms in Figure~\ref{expected} show distributions of relative angle between the scattering planes
for data selected from the region: $(\theta_{1} - 81.66^{\circ})^2 + (\theta_{2} - 81.66^{\circ})^2 \leq R^2$, choosing two exemplary radii of $R~=~10^\circ$ and $R~=~30^\circ$, where the high visibility is expected. 
Figure~\ref{expected} compares results obtained for the case of (i) the ideal detector with 100\% efficiency and infinitely good angular resolution for $R~=~10^\circ$ (black solid line) and $R~=~30^\circ$ (red solid line) with (ii) distribution of the relative angle between the scattering planes reconstructed based on the interaction positions simulated in the detector for $R~=~30^\circ$ (red dashed line) and with an additional condition that the interaction points should be more distant than $12$~cm (red dotted line). The last condition is applied in order to ensure good angular resolution ($\sim 2^\circ$) and good  selection power for primary and secondary interactions. The expected interaction time resolution of $100$~ps~\cite{PMB2016} corresponds to about $4.2$~cm resolution for the measurement of the distance between the interaction points. Thus the requirement of $12$~cm separation between interaction points should allow for assignments of primary and secondary interaction at the purity of $3\sigma$.

Relative angle between the scattering planes $\varphi$ is an estimator of the relative angle between the polarization directions of the registered photons. As discussed in the introduction, the distribution of this angle indicates the uncertainty (resolution function) of determining relative angle between polarization directions on an event by event basis. The shape of these resolution functions (shown with solid lines in Fig.~\ref{expected} for the ideal detector in two chosen regions of high visibility) are determined by the nature of the Compton scattering (Klein-Nishina formula). Comparing red and black distributions one observes, as expected, that the smaller is the area around the most optimal scattering angle more enhanced is the maximum around $\varphi$~=~90$^\circ$. The additional modification of these distributions are due to the angular resolution and the specific geometry of the detector. Red-dashed line indicates histogram after requiring that the distance between the interactions is larger than $d~>~12$~cm and the blue-dotted histogram shows final expected results assuming in addition that the energy loss in the scintillators for each interaction must be larger than $50$~keV (this requirement emulates the electronic thresholds of the J-PET device).

\begin{figure}[h]
  \includegraphics[width=0.5\textwidth]{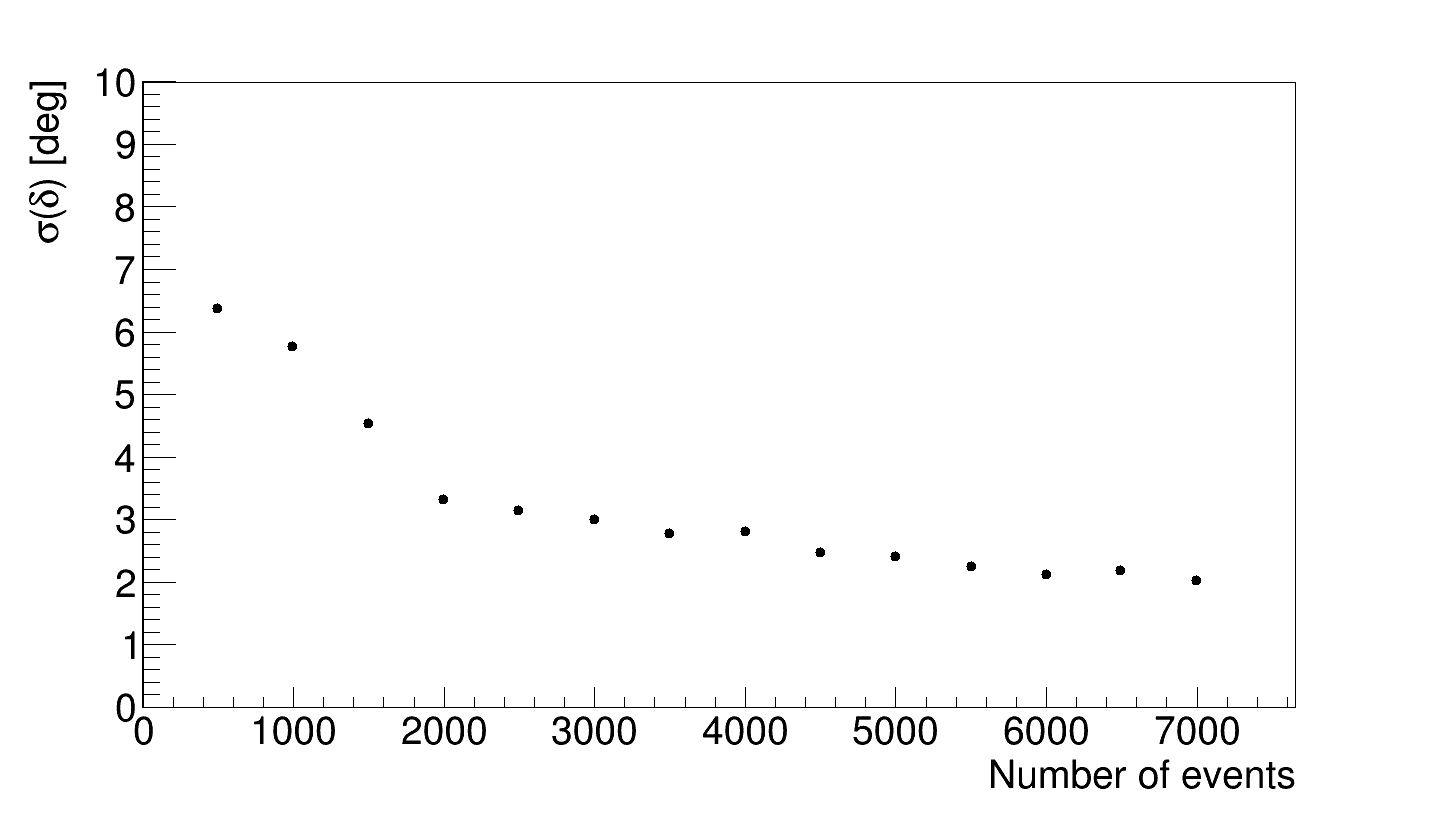} 
\caption{
Uncertainty of the determination of the relative angle between the photons polarization directions ($\delta$ parameter defined in Eq.~\ref{cosine}) as a function of number of measured events for the J-PET setup shown in Fig~\ref{DetectorSchemaSingle}.
\label{4LayUncert}
}
\end{figure}

Results presented in Fig.~\ref{fig:Efficiency-Theta1-vs-Theta2} and Fig.~\ref{expected} indicate that J-PET covers the full angular phase space with no holes in the efficiency map. The efficiency for the measurement of relative angle $\varphi$ 
is smooth and nearly constant. These features enable reliable corrections of the measured $\varphi$ distributions for the efficiency.  

As an estimator of the average relative angle between polarization direction of the back-to-back photons, a parameter $\delta$ may be used, which can be determined by fitting equation~(\ref{cosine}) to the efficiency corrected distribution of angle $\varphi$. Uncertainty of the parameter $\delta$  is decreasing with the number of registered events.  Fig.~\ref{4LayUncert} indicates that the uncertainty (standard deviation) of the average relative angle between the polarization of the back-to-back photons is equal to few degrees already for a sample of about $5000$ registered event.

Last but not least let us comment on the theoretically predicted entanglement of the two gammas. Taking latter into account would result in a double as big total visibility, then observed in Fig.~\ref{TheoSim} and Fig.~\ref{expected}. This factor stems from the fact that photons are bosons and obey the Bose-Einstein symmetry.

\section{Summary and perspectives}
\label{section5}
Measurements of optical photon's polarization have a long successful history in physics, constituting the basis for investigations of phenomena connected with quantum entanglement of photons such as quantum teleportation or quantum cryptography. 
In this article we explored the possibility of estimating the polarization of high energetic photons originating from the decays of positronium atoms with the novel technology of the J-PET detector. For the first time, polarization studies become possible in this energy regime and, by that, studies of photonic entanglement five orders of magnitude beyond the optical wavelength regime.  

J-PET is the first PET tomograph built from plastic scintillators in which annihilation photons are measured via  Compton scattering. 
We have shown that the polarization of photon, at the moment when it scatters on electron via Compton effect, can be estimated on an event by event basis. We have studied possibilities of estimating the photon's linear polarization at the moment of its interaction with the electron by the cross product of the momentum vectors 
$\hat{k} \times \hat{k'}$ before and after the scattering. Based on this definition it was shown that in case of two back-to-back photons, the relative angle between their polarization directions may be estimated by the relative angle between their scattering planes.

Our simulations indicated that, for the ideal detector, due to the nature of the Compton effect, 
the resolution (visibility of the polarization) strongly depends on the scattering angle, achieving a standard deviation of $\sigma=40^\circ$ for 
$\theta\approx 82^\circ$, and worsening towards smaller and larger scatterings angles. For forward and backward scatterings the measurement of the polarization via Compton effect becomes impossible.
Furthermore, simulations performed with the GATE programming package~\cite{Santin2003,Jan2004,Jan2011}, including the geometry and material composition of the J-PET detector showed that the efficiency for the measurement of the polarization of $511$~keV photons originating from the positronium decay is smooth and relatively high. 
In the region of high visibility (circle with the radius of R~=~30$^\circ$ around $\theta_1~=~\theta_2~=~ 81.66^\circ$ - the highest visibility), the efficiency of the J-PET detector updated with a fourth layer (Fig.~\ref{DetectorSchemaSingle}) amounts to about 0.2\%. However, due to the small cross section in this angular range (see Fig.~\ref{fig:Theory-Theta1-vs-Theta2}) and the additional selection criteria such as that the distance between interaction is larger than $12$cm and the energy deposit for each interaction is larger than $50$~keV the total detection efficiency amounts to about $10^{-6}$. This efficiency was calculated as the ratio of number of events for which both two primary and Compton scattered photons were registered in the region of high visibility $R~=~30^\circ$ to the overall number of simulated para-positronium decays (Fig.~\ref{fig:Theory-Theta1-vs-Theta2}).
Thus assuming that for the four-layer J-PET (Fig.~\ref{DetectorSchemaSingle}) the final total detection and selection efficiency will be equal to $10^{-6}$, 
we expect about ten events of interest (Fig.~\ref{DetectorSchemaSingle}) per second when using the sodium $^{22}Na$ source with activity of $10 \times 10^6$~MBq surrounded with the XAD4 porous polymer~\cite{JasinskaActa-2016}.  This will in practice allow for obtaining statistics from million of events within a few days of measurements. 

Finally, we have shown that the angular resolution obtainable with the J-PET detector, for the determination of the relative mean angle between the linear polarization of the back-to-back propagating annihilation photons is equal to about $\sigma(\delta) \approx 2^\circ$ for samples of 5000 or more collected events. 

The results are encouraging and show that it is feasible to perform measurements of the quantum entanglement of photons from positronium annihilation~\cite{Beatrix-Science-Report2017,beatrix-arXiv:1807.04934}
with the J-PET detector. In particular, determination of the polarization on an event-by-event basis will enable, for the first time, tests of entanglement in the polarization degrees of freedom of the three photons resulting from the decay of the ortho-positronium~\cite{Beatrix-Science-Report2017} as well as tests of the discrete symmetries, parity $P$, time reversal $T$ and charge-conjugation--parity $CP$, via 
 operators $\epsilon_i \cdot \vec{k_j}$, where the indices $i,j=1,2,3$ refer to the labeled photons from the ortho-positronium decays. 
Such discrete symmetries tests, carried out with the J-PET detector~\cite{ACTA2016}, are complementary to so far performed experiments where the operators are constructed from spin observables ($\vec{S}$) of ortho-positronium and photon's momentum vectors~\cite{Yamazaki2010,Vetter2003}.
Violation of the $T$ or the $CP$ invariance in purely leptonic systems has never been seen so far~\cite{Kosteleck2011}. The experimental search is limited by effects due to the photon-photon interactions expected to mimic discrete symmetry violations at the level of 10$^{-9}$~\cite{Arbic1988,Bernreuther1988}~\footnote{The contribution from weak interaction to the positronium decays can be neglected with respect to photon-photon interaction. For example the branching ratio of $C$ violating Ps decays due to the Z and W bosons is expected at the level of $10^{-27}$~\cite{Bernreuther1981}  and $10^{-77}$~\cite{Pokraka2017}, respectively.}. Therefore, there is still a range of about six orders of magnitudes with respect to the present experimental limits (currently experimental upper limits for $T$, $CP$ and $CPT$ violations are at the level of 10$^{-3}$~\cite{Yamazaki2010,Vetter2003}) where phenomena beyond the Standard Model can be sought for. The J-PET detector offers therefore a new experimental methodology.

\textbf{Acknowledgement}
The authors acknowledge technical and administrative support by A. Heczko, M. Kajetanowicz and W. Migda\l{}. This work was supported by The Polish National Center for Research
and Development through grant INNOTECH-K1/IN1/64/159174/NCBR/12, the
Foundation for Polish Science through the MPD and TEAM/2017-4/39 programmes, the National Science Centre of Poland through grants no.\
2016/21/B/ST2/01222,\linebreak[3] 2017/25/N/NZ1/00861,
the Ministry for Science and Higher Education through grants no. 6673/IA/SP/2016,
7150/E-338/SPUB/2017/1,  \\  7150/E-338/M/2017 and 7150/E-338/M/2018,
and the Austrian Science Fund FWF-P26783.

\end{document}